\documentclass[journal]{IEEEtran}

\usepackage{graphicx}
\usepackage{amsmath}
\usepackage{amssymb}
\usepackage{marvosym}
\usepackage{epstopdf}

\newcommand{\fatJ}{\mathbf{J}}
\newcommand{\fatE}{\mathbf{E}}

\newcommand{\fatD}{\mathbf{D}}

\begin{document}

\title{Finite element analysis of neuronal electric fields: the effect of heterogeneous resistivity}

\author{Pavol Bauer,
  Sanja Mikulovic,       
  Stefan Engblom,
  Katarina E Le\~{a}o,
  Frank Rattay,
  Richardson N Le\~{a}o
\thanks{Pavol Bauer and Stefan Engblom are with
  Division of Scientific Computing, Department of Information Technology,
  Uppsala University,  SE-751 05 Uppsala, Sweden, e-mail: pavol.bauer@it.uu.se}%
\thanks{Sanja Mikulovic, Katarina E Le\~{a}o and Richardson N Le\~{a}o are with
  Department of Neuroscience, Uppsala University
  Box 593, 751 24 Uppsala, Sweden, e-mail: richardson.leao@uppsala.uu.se}%
\thanks{Sanja Mikulovic and Frank Rattay are with
  Institute for Analysis and Scientific Computing, 
  Vienna University of Technology
  Vienna, Austria}%
\thanks{Katarina E Le\~{a}o and Richardson N Le\~{a}o are with
  Brain Institute, Federal University of Rio Grande de Norte 
  Natal-RN, Brazil}}

\date{Received: date / Accepted: date}

\maketitle

\begin{abstract}
  Simulation of extracellular fields is one of the substantial methods
  used in the area of computational neuroscience.  Its most common
  usage is validation of experimental methods as EEG and extracellular
  spike recordings or modeling of physiological phenomena which can
  not be easily determined empirically.  Continuous experimental work
  has been re-raising the importance of polarization effects between
  neuronal structures to neuronal communication.  As this effects rely on 
  very small potential changes, better modeling methods are necessary to quantify the weak
  electrical fields in the microscopic scale in a more realistic way.

  An important factor of influence on local field effects in the
  hippocampal formation is the heterogeneous resistivity of
  extracellular tissue.  The vast majority of modeling studies
  consider the extracellular space to be homogeneous while
  experimentally, it has been shown that the stratum pyramidale has
  two times higher resistivity than other hippocampal layers.  Common
  simulation methods for extracellular electrical fields based on the
  point source approximation are bound to describe the resistance of
  the space with a single scalar. We propose that models should be
  based on the space- and time-dependent Maxwell equations (Partial
  Differential Equations, PDEs) in order to account for heterogeneous
  properties of the extracellular space and specific arrangements of
  neurons in dense hippocampal layers.

  To demonstrate the influence of heterogeneous extracellular
  resistivity and neuronal spatial orientation on modeling results, we
  combine solutions of classical compartment models with
  spatiotemporal PDEs solved by the Finite Element Method (FEM). With
  the help of these methods, we show that the inclusion of
  heterogeneous resistivity has a substantial impact on voltages in
  close proximity to emitting hippocampal neurons, substantially
  increasing the change in extracellular potentials compared to the homogeneous
  variant.
\end{abstract}


\section{Introduction}
\label{intro}

Numerous computational studies have investigated time-varying currents
in homogeneous extracellular space \cite{mcintyre_excitation_1999,
  bedard_modeling_2004, bedard_model_2006} as well as the role of neuronal
morphology in uniform electric field stimulation
\cite{chan_effects_1988, radman_spike_2007,
  radman_role_2009}. However, most studies use models of the
extracellular milieu that may not be accurately applied when modeling
brain structures with highly diverse extracellular resistivity and
neuronal arrangement.

As emphasized by Lopez-Aguado and Bokil
\cite{lopez-aguado_activity-dependent_2001, bokil_ephaptic_2001} it
has been traditionally neglected that currents propagate in all
directions in an extracellular medium and that inward and outward
currents originate from tissue regions having large resistivity
differences. A usual argument for this approach is that resistivity
influences extracellular electrical fields minimally and that
extracellular space can be assumed as homogeneous in current-source
density (CSD). However, extracellular non-homogeneous resistivity has
been shown experimentally in several regions of the brain, for
example, in the hippocampus and the cerebellum
\cite{okada_origin_1994}. Additionally, tissue swelling has been
observed after intense neural activity that in turn could lead to an
increase in extracellular resistivity \cite{autere_synaptic_1999}.
Previous studies \cite{mcintyre_excitation_1999, gold_origin_2006,
  gold_using_2007, anastassiou_effect_2010} have attempted to examine
how extracellular electrical fields affects neuronal
activity although with the help of quasi-static approximation and an
assumed homogeneous tissue resistivity.

To quantify the effect of extracellular heterogeneous resistivity and
neuronal spatial distribution on strength of neural fields, we are
proposing a simple modeling pathway to couple compartment-based neural
models with the COMSOL Multiphysics simulation environment. Here, we
solve time-dependent Maxwell's equations using the FEM to analyze the
change of electrical fields as it occurs in the extracellular space
surrounding neurons.

Our results indicate that inhomogeneous resistivity of the
extracellular milieu significantly influences the change of
extracellular potentials (EPs) in the hippocampus. By computing the
resulting voltage change due to outgoing transmembrane currents of an
exemplar CA1 pyramidal cell model in homogeneous and inhomogeneous
extracellular resistivity, we observed a maximal difference in EP
change of 60\% in the hippocampal pyramidal layer. Furthermore, the
here proposed method offers the possibility to efficiently simulate
the effects of superimposed extracellular potentials created by
neurons in divergent positions relative to each other. We will also
discuss advantages and drawbacks of the used method and propose
alternatives and possible improvements towards more realistic modeling
of electrical fields of the brain.


\section{Materials and Methods}
\label{sec:1}
\subsection{Pyramidal neuron model}

To demonstrate the effect of trans-membrane currents on the effect of
extracellular fields, a Hodgkin-Huxley (HH)-like hippocampal CA1
pyramidal cells was adapted from \cite{pinsky_intrinsic_1994} with the
addition of a current $I_{h}$ from
\cite{leao_hyperpolarization-activated_2005} and a current $I_{m}$
from \cite{leao_kv7/kcnq_2009}.

$I_{h}$ current:
\begin{equation}
  u_{\infty} = [1+\exp((V+76)/7)]^{-1},
\end{equation}
\begin{equation}
\begin{aligned}
  \tau_{u} &= 10^{4}/[237 \exp((V+50)/12)+ \\
    &\phantom{= 10^{4}/[} 17\exp(-(V+50)/25)]+0.6,
\end{aligned}
\end{equation}
\begin{equation}
  I_{h}(t,V) = g_{h} u(t,V) \, (V-E_{h}).
\end{equation}

$I_{m}$ current:
\begin{equation}
  s_{\infty} = [1+\exp(-(V+22.53)/10)]^{-1},
\end{equation}
\begin{equation}
\begin{aligned}
  \tau_{s} &= 4135.7/[164.64 \exp ((V-0.05) \cdot 0.12)+ \\
    &\phantom{= 4135.7/[}0.33 \exp(-(V-0.05)/10)]+35.66,
\end{aligned}
\end{equation}
\begin{equation}
  I_{m}(t,V) = g_{m} s(t,V) \, (V-E_{K}).
\end{equation}
Our model contained one somatic (r = 10 $\mu m$), 20 dendritic (r =
1--3 $\mu m$, l = 5 $\mu m$), two axon initial segments, $AIS_{1}$ (r =
0.5 $\mu m$, l = 60 $\mu m$ ), $AIS_{2}$ (r = 0.5 $\mu m$, l = 60
$\mu m$), and finally 32 axonal compartments (r = 0.5 $\mu m$, l = 5
$\mu m$), see Figure~\ref{fig:1}. Note that here $AIS_{1}$ refers to
the part of AIS from 0 to 60 $\mu m$, and $AIS_{2}$ from 60 to 120
$\mu m$ from the cell body. Additionally, $AIS_{2}$ compartments
contain high $Na^{+}$ channel density \cite{palmer_site_2006}. 20
dendritic compartments are used for the dendritic tree, while
additional compartments are included in the branching analysis. The
$Na^{+}$ channel density was varied between 986 and 2943 $\frac{\mu
  A}{cm^2}$ (corresponding to $AIS_{2}$). The following conductance
values were used: $g_{Na}$ = 8, $g_{kdr}$ = 5, $g_{m}$ = 1 $mS/cm^2$
for the soma; $g_{Ca}$ = 10, $g_{K_C}$ = 15, $g_{K_AHP}$ = 0.8,
$g_{O}$ = 0.625, $g_{Na}$ = 0.07, $g_{h}$ = 0.2 $mS/cm^2$ for the
dendrite; $g_Na$ = 50, $g_{kdr}$ = 10, $g_{m}$ = 10 $mS/cm^2$ for the
$AIS_{2}$ and $g_{Na}$ = 9, $g_{kdr}$ = 10, $g_{m}$ = 10 $mS/cm^2$ for
the rest of the axon compartments. The leakage conductance was set to
0.1 for all of the compartments. Finally, the equilibrium potentials
were set to $E_{Na}$ = 60 mV, $E_{K}$ = -85 mV, $E_{h}$ = -43 mV and
$E_{leakage}$ = -65 mV.

\subsection{Creation of morphology and model coupling}

\begin{figure}
  \includegraphics[width=0.5\textwidth]{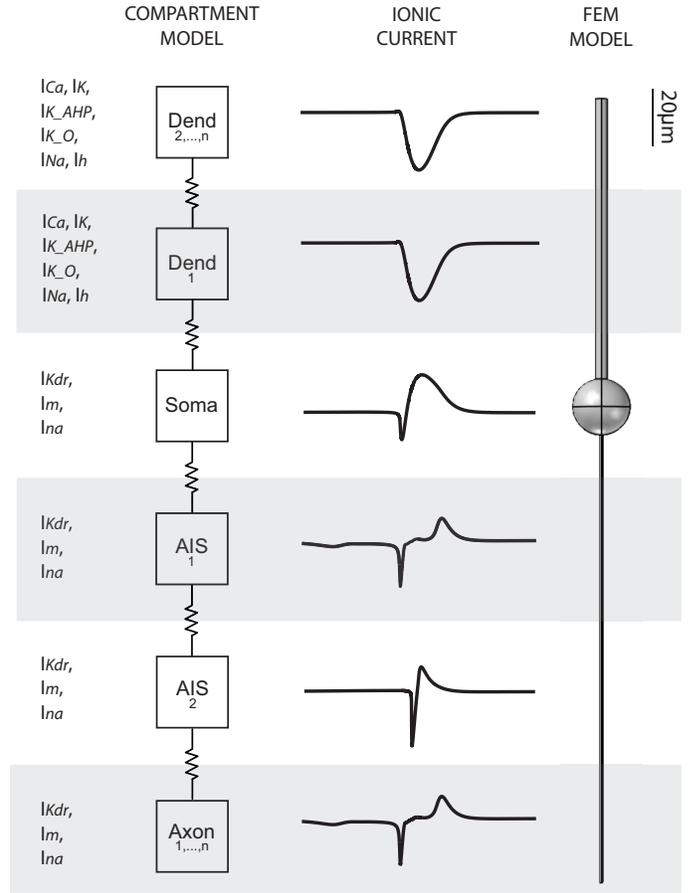}
  \caption{\textbf {Mapping trans-membrane currents from compartmental
      model to FEM boundary mesh}. A compartmental model containing 20
    dendritic, one soma, two axon initial segments and 32 axonal
    compartments. Ionic currents for each compartment are shown on the
    left. The compartmental model outputs $I(t)$ which is used as
    current boundary sources $Q_j(t)$ in the corresponding
    three-dimensional volumes in the FEM model.}
  \label{fig:1}
\end{figure}

The three-dimensional neuronal geometry was constructed in COMSOL
Multiphysics 4.3 with the help of the interface to MATLAB
(``LiveLink'') by morphological additions and boolean unification of
simple geometric volumes. As we aim to represent the 3D-morphology as
an exact counterpart of the compartmental model, each section is
recreated as a cylinder with the same length and diameter as in the
compartmental definition. If the cylinders were added on top of each
other with no change of the rotation vector, as shown in Figure
\ref{fig:1}, no joining geometrical primitives were added in between
them. If the rotation differs, as for example in Figure \ref{fig:4}, a
sphere is added in between the cylinders, followed by a removal of the
interior boundaries of both the cylinders and the sphere. The reason
for constructing the geometry in this way is that the mesh engine
otherwise respects the internal boundaries such that the resulting
mesh becomes unnecessary complex. In some cases the occurrence of
multiple internal boundaries can even make the meshing procedure
abort.

Ionic currents of single neuronal compartments model were determined
by solving Hodgkin-Huxley Ordinary Differential Equations (HH ODEs)
specified in equations 1 to 6 using a Runge-Kutta algorithm from the
MATLAB ODE-suite (Mathworks) with a constant step-size $T$. The sum of
all currents per compartment for each time step of the simulation
$I_0$ is afterwards normalized to an absolute value of units $A/m^2$
and stored in a matrix $I$ of size $N_{compartments} \times
N_{timesteps}$.

Next, by using the interface to COMSOL, each row of the matrix $I$ was
mapped adequately to the corresponding cylindrical domain as a boundary
current source $Q_j(t)$.  Note that we hereby assume that
transmembrane currents are the only cause of change of extracellular
potential, which is surely not the case in a real neuron, as for
example synaptic calcium-mediated currents are suspected to
contribute to a large fraction of the extracellular signature
\cite{buszaki_review_2012}.

The extracellular volume was modeled by cylindrical objects covering
the neuronal morphology with either homogeneous (0.3~S/m) or
heterogeneous resistivity as shown in Figure~\ref{fig:2}A. In this case
we constructed the extracellular volume by taking the union of
cylindrical objects with increasing resistivity in the $y$-axis,
according to \cite{lopez-aguado_activity-dependent_2001}.  It is
assumed that the conductivity of extracellular tissue is
frequency-independent in the used range of neural activity
(10--100~Hz) \cite{linden_intrinsic_2010}.

\subsection{Electrostatic formulation and the Finite Element Method}

To simulate non-homogeneous distributions of electrical fields
produced by single neurons and neuronal networks, we used an
electrostatic formulation of Maxwell's equations discretized with
finite elements. Chiefly, in the finite element approach, Maxwell's
equations are solved by discretizing the incorporated volumes (in this
case the neuronal compartments) into finite tetrahedral volume
elements \cite{JinEM_FEM}. We seek the \emph{electric field
  intensity} $\fatE$ in terms of the \emph{electric scalar potential}
$V$,
\begin{equation}
  \label{eq:gauge}
  \fatE = -\nabla V.
\end{equation}
The relevant dynamic form of the continuity equation with current
sources $Q_{j}$ is given by
\begin{equation}
  \label{eq:cont}
  \nabla \cdot \fatJ = -\frac{\partial \rho}{\partial t}+Q_{j},
\end{equation}
with $\fatJ$ and $\rho$ the \emph{current density} and \emph{electric
  charge density}, respectively. Further constitutive relations
include
\begin{equation}
  \label{eq:const1}
  \fatD = \varepsilon_{0} \varepsilon_{r} \fatE,
\end{equation}
and \emph{Ohm's law}
\begin{equation}
  \label{eq:const2}
  \fatJ = \sigma \fatE,
\end{equation}
in which $\fatD$ denotes the \emph{electric flux density}. Finally,
\emph{Gauss' law} states that
\begin{equation}
  \nabla \cdot \fatD = \rho.
\end{equation}
Upon taking the divergence of \eqref{eq:const2} and using the
continuity equation \eqref{eq:cont} we get
\begin{equation}
  \nabla \cdot \sigma \fatE = 
  -\frac{\partial \rho}{\partial t}+Q_{j}.
\end{equation}
Rewriting the electric charge density using Gauss' law together with
the constitutive relation \eqref{eq:const1} and finally applying the
gauge condition \eqref{eq:gauge} twice we arrive at the time-dependent
potential formulation
\begin{equation}
  \label{eq:form}
  -\nabla \cdot \left(\sigma \nabla V+
  \varepsilon_{0}\varepsilon_{r} \frac{\partial}{\partial t}
  \nabla V \right) = Q_{j}.
\end{equation}
This is the formulation used in COMSOL Multiphysics
\cite{ComsolACDC}. The values for the electric conductance $\sigma$
and the relative permittivity $\varepsilon_{r}$ were obtained from
\cite{bedard_modeling_2004}. The source currents $Q_{j}$ were
computed from the compartmental model as described above.

The formulation \eqref{eq:form} is efficiently solved by COMSOL's
``Time discrete solver'', which is based on the observation that the
variable $W := \Delta V$ satisfies a simple ODE. Solving for $W$ in an
independent manner up to time $t$, it is then straightforward to solve
a single static PDE to arrive at the potential $V$ itself.

As for boundary conditions we took homogeneous Neumann conditions
(electric isolation) everywhere except for in a single point which we
choosed to be ground ($V = 0$). In all our simulations this point was
placed at the axis of rotation of the enclosing cylindrical
extracellular space, and underneath the neuronal geometry. This
procedure ensures that the formulation has a unique
solution (it is otherwise only specified up to a constant).

A tetrahedral mesh was applied to discretize space (using the
``finest'' mesh setting; resolution of curvature 0.2, resolution of
narrow regions 0.8). The simulations were verified against coarser
mesh settings in order to ensure a practically converged solution. As
a final note, in the Time discrete solver the time step was set to
the same step size $T$ as used in the ODE-based solution of
the Hodgkin-Huxley equations, thus ensuring a correct transition
between both simulation environments.

\section{Results}

\subsection{The effect of heterogeneous extracellular fields}

\begin{figure}
\includegraphics[width=0.5\textwidth]{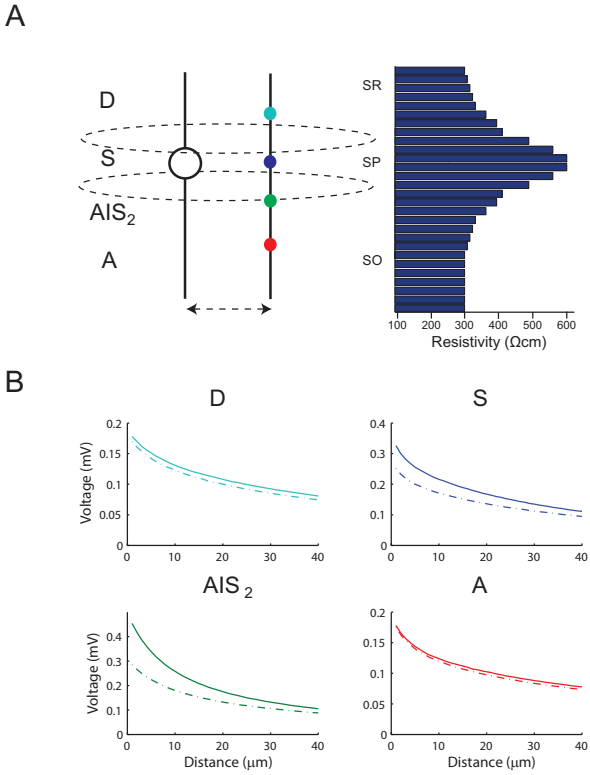}
\caption{\textbf{Heterogeneous extracellular resistivity modulates the
    strength of polarization effects in proximity of the firing
    neuron}. A, Schematic representation of a model neuron impact on
  defined point probes. Point probes act as receivers and are being
  shifted away from the active neuron along the $x$-axis
  \textit{(left)}.  Non-homogeneous resistivity distribution of
  hippocampal regions \textit{(right)}. B, Relationship between
  membrane potential at different point probes affected by dendritic (\textit{D}),
  somatic (\textit{S}), axon initial segment ($AIS_{2}$) and axon terminal (\textit{A}) neuronal compartments (color
  coded as in \textit{A}) for distance varying between 0 and 80 $\mu
  m$. \textit{Dashed:} the resulting voltage change assuming
  homogenous extracellular resistivity (= 350 $\Omega$ cm), while
  \textit{solid} represents the result of assuming a heterogenous
  resistivity.}
\label{fig:2}
\end{figure}

In order to examine the influence of heterogeneous extracellular
space, we first constructed a three-dimensional active neuron model in
an homogeneous and heterogeneous extracellular milieu (Figure
\ref{fig:2}).  Neurons in the hippocampus have an intricate spatial
orientation that propitiates strong field potentials: high density of
pyramidal cell dendrites running in parallel in the stratum radiatum
(SR), densely packed pyramidal cell somas in the stratum pyramidale
(SP) while pyramidal cell axons run almost in parallel or crossing
each other in SP or stratum oriens (SO).

We measured the voltage on four defined point probes placed parallel to the dendrites, soma, $AIS_{2}$ and axon
terminal compartments during the peak of an action potential (AP) of 80mV, by
varying the distance between active neuron and point probes from 1 $\mu m$ to 80 $\mu m$ (Figure \ref{fig:2}A).

In the first set of simulations, we analyzed the effect of the
aforementioned four neuronal regions on the defined point probes
assuming a widely accepted homogeneous resistivity of 350 $\Omega$ cm
\cite{bedard_modeling_2004, bedard_model_2006}. Note that here applied
boundary currents of three-dimensional neuronal compartments
correspond to the peak transmembrane current during an AP. By doing
so, the peak voltage of 0.25 mV was obtained in the point
probe parallel to the $AIS_{2}$ compartment, followed by soma,
dendrite and axon terminal.  Consequently, the heterogeneous case was
examined by placing the neuron and point probes into a heterogeneous
extracellular space representing hippocampal spatial order, where
resistivity values for different strata were obtained from
\cite{lopez-aguado_activity-dependent_2001} and shown in Figure
\ref{fig:2}A.  The active neuron was positioned in the center of SP
and point probes were moved along the x axis of the extracellular
space. In the case of non-homogeneous extracellular resistivity
(Figure \ref{fig:2}C) the largest voltage change was measured
in point probes parallel to $AIS_{2}$, although the values in point probes placed in SP were 60\% larger 
in close distance than in the homogeneous extracellular space
scenario and 28\% higher considering the spatial mean over the total distance.
The point probes placed in SO and SR were as well affected by the higher restivity of the 
pyramidal layer, showing an average increase of 4\% in parallel to the axon and 7\% in parallel to the dendrite.
The appreciable difference between the voltage changes
suggests that non-homogeneous distribution of resistivity is an
important aspect of extracellular field effects.

\begin{figure}
  \includegraphics[width=0.5\textwidth]{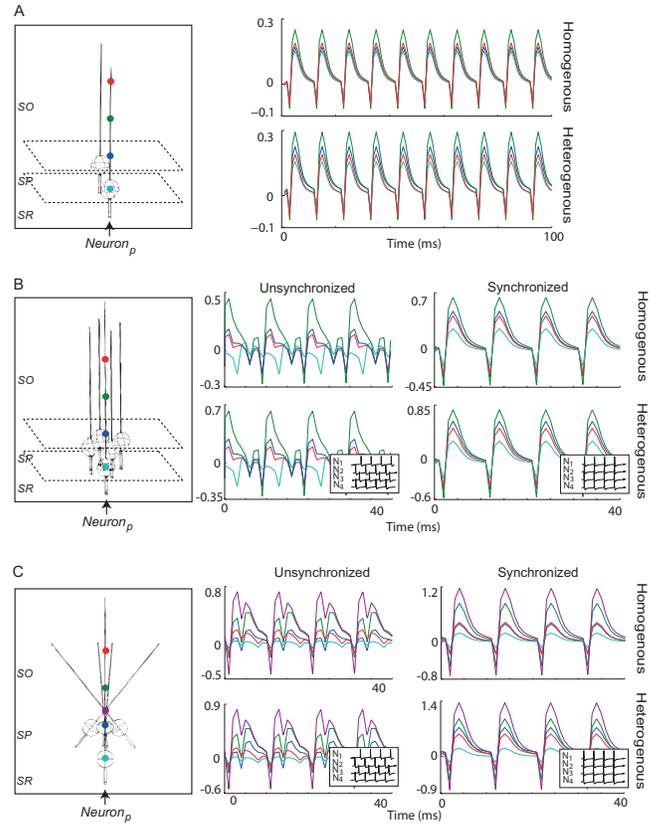}
  \caption{\textbf {Analysis of extracellular fields in dependence of
      spike timing and spatial orientation}.  A. Potential change
    calculated at different regions (colored circles) of a passive
    neuron (\textit{$Neuron_{p}$}) in response to EPs triggered from
    action potentials in the initial phase of an active neuron. Lines
    in the \textit{right} panel are color coded as the circles in the
    \textit{left} panel.  Somas were placed at the stratum pyramidale
    and extracellular resistivity followed the same distribution as
    shown in Figure 2C.  Upper trace of the right panel shows
    potential changes in homogeneous, while lower trace in
    heterogeneous extracellular space. B, Same as in \textit{A} but
    the \textit{$Neuron_{p}$} surrounded by four active
    neurons. Potential change at different regions of
    \textit{$Neuron_{p}$} when four active neurons fire asynchronously
    (middle panel) and synchronously (right panel). \textit{Insets} on
    the middle and right panels show the firing of active neurons). C,
    Same as in \textit{B}, but the active neurons where oblique (but
    not intersecting, nearest interaxonal distance = 2 $\mu m$) at the
    position of $AIS_{2}$ to \textit{$Neuron_{p}$}.}
  \label{fig:3}
\end{figure}

\begin{figure}
  \includegraphics[width=0.5\textwidth]{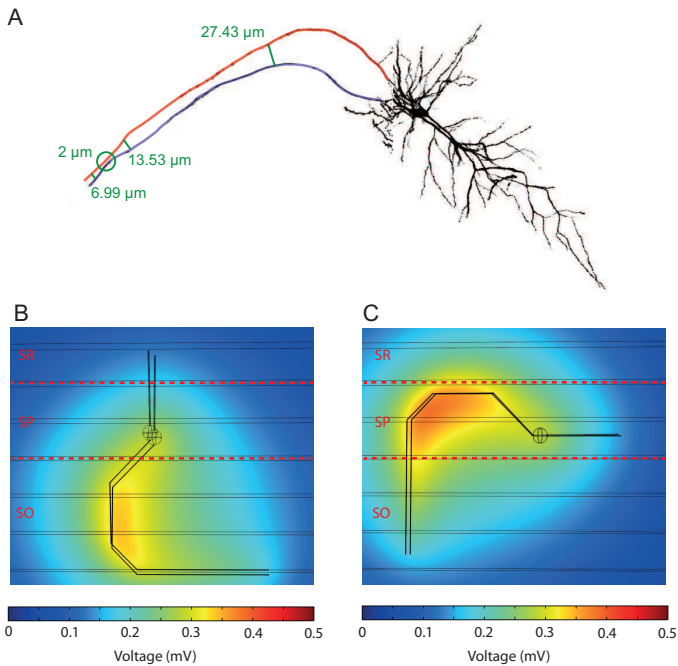}
  \caption{\textbf {Modeling the effect of bended axons}. A, Confocal
    image of two neurons in the hippocampus filled with neurobiotin for
    \textit{post hoc} spatial analysis showing two parallel neurons
    with bent axons situated at a minimal distance of 2 $\mu m$. B, A
    simplified representation of both neurons when the nearest
    interaxonal region was confined in stratum oriens. In this case,
    an action potential in the active neuron generates a voltage
    transient of 0.25 mV at the passive neuron soma and 0.35 mV at the
    neuron $AIS_{2}$. The figure shows the extracellular voltage
    distribution at the peak amplitude of the active cell action
    potential. C, Same as \textit{B}, but axon initial segments were
    confined at the stratum pyramidale (higher resistivity). In this
    case, a 0.25 mV and 0.45 mV voltage transients were observed at
    the passive neuron soma and axon, respectively.}
\label{fig:4}
\end{figure}

\subsection{Analysis in respect to spatial orientation and neural morphology}

According to the superposition principle electrical fields (EFs) are
produced by summation of single neuronal activity. Thus, the mutual
interaction between neuron and field is strongly modulated by the
spatial orientation of neuronal assemblies
\cite{froehlich_endogenous_2010}. To analyze EFs considering a
realistic spatial order, we began simulating two neurons with parallel
axons (Figure \ref{fig:3}A) where $Neuron_a$ represents an active
structure with boundary current source $Q_j$ mapped to its surface
while $Neuron_p$ is a passive measurement structure. We measured
$V_{e}$ at four points along the axon of \textit{$Neuron_{p}$} and
detected the maximal voltage amplitude of 0.28 mV. Note that here
$V_{e}$ corresponds to extracellular voltage measured on the cell
membrane. Next we tested the influence of four neighboring cells with
parallel axons on the \textit{$Neuron_{p}$} (Figure \ref{fig:3}B). In
this case, having four active neuronal neighbors (distance between the
axons =10 $\mu m$) firing non-synchronously, a maximal voltage of 0.49
mV in homogeneous and 0.68 mV in heterogeneous extracellular medium of
\textit{$Neuron_{p}$} was computed. Synchronous firing of adjacent
neurons produced a peak voltage of 0.62 mV in homogeneous and 0.84 mV
in heterogeneous medium of $AIS_{2}$. When axons crossed each other at
the $AIS_{2}$ region (Figure \ref{fig:3}C), we observed a maximal
voltage amplitude of 0.89 mV in the $AIS_{2}$ of \textit{$Axon_{p}$}
during asynchronous activity of the neighboring cells, whereas
synchronous activity produced 1.41 mV in heterogeneous medium. These
results suggest that for analyzed neuronal arrangement soma and
$AIS_{2}$ have considerable influence on the strength of neural fields
for small distances ($\leq$ 10 $\mu m$) and that values in
heterogeneous extracellular space are in mean $\sim$ 25\% larger
than in homogeneous extracellular medium when averaged over time. 

Furthermore, we simulated the influence of pyramidal neuron position
within different hippocampal layers on the strength of neural fields
(Figure \ref{fig:4}).  We positioned somas in SP with the smallest
distance between the two axon initial segments confined to SO. This
geometrical arrangement was motivated by confocal images of two
proximal pyramidal neurons located parallel to each other with an
axonal bend starting about 50 $\mu m$ behind the soma (Figure
\ref{fig:4}A).  As known from previous analytical studies
\cite{tranchina_model_1986} the bend of axonal structures generally
amplifies endogenous fields around neurons.  We were able to confirm
this effect in our simulation as we simulated the field effect
resulting from trans-membrane currents of an active neuron measured on
the parallel neuron.  In this scenario, peak voltage of 0.35 mV was
calculated at the position of $AIS_{2}$ while the smallest voltages
were registered at the passive neurons between the dendrites (0.12 mV)
in SR and axon terminals (0.16 mV). If $AIS_{2}$ were confined to SP,
the potential computed at the passive neuron was 25\% larger than in
the case in which $AIS_{2}$ were positioned in the SO (Figure
\ref{fig:4}B).


\section{Discussion}

In this work, we use the Finite Element Method coupled to the
HH-equations to simulate how neuronal geometry, arrangement and
heterogeneous extracellular properties affect the strengths of neural
fields. We first show that there is a difference in voltage transients
produced by firing of neighboring cells if the extracellular space is
considered to be homogeneous or heterogeneous. Additionally, we
demonstrate that the spatial orientation of specific cellular
compartments is an important determinant of the strength of neural
fields.

In our computations, the highest change of extracellular potential
arises in the pyramidal layer, in proximity to the $AIS_{2}$
compartment. Action potentials are generally initiated in the AIS due
to a higher density of voltage-gated $Na^{+}$ channels
\cite{palmer_site_2006, bean_action_2007, kole_is_2008}, which is
reflected by the parameter settings for this compartment in the
occupied neuronal model. Additionally, the spatial arrangement of
hippocampal pyramidal neurons also propitiates the proximity of
$AIS_{2}$ in both SP and SO (see Figure~\ref{fig:4},
\cite{ramon_y_cajal_histologie_1911}).

Holt and Koch \cite{holt_electrical_1999} showed that interactions
near cell bodies are more important than interactions between axons by
using standard one-dimensional cable theory and volume conductor
theory. Another study \cite{traub_simulation_1985} reported a 4.5 mV
change (in the AIS) caused by extracellular interactions, a change more
than 4$\times$ larger than what we have found in our simulations of 
heterogeneous extracellular media. Additionally, by applying analytical methods,
Bokil et al \cite{bokil_ephaptic_2001} have shown that, in the
olfactory system, an action potential of 100 mV amplitude in one axon
could produce depolarization in other axons in the bundle sufficient
to initiate an action potential. However, using FEM simulations it was
not possible to trigger spikes in neurons solely by electrical fields
mediated by in- and outflow of trans-membrane currents, in agreement
with the work by Traub and colleagues (in which maximal voltages in a
sink axon during synchronized activity of four neighboring neurons is $ \approx 1.2 mV$ ).

Hence, simulations relying on the point source approximation to
describe hippocampal neural fields may be distorted as the potential
change is more than 28\% greater in stratum pyramidale and in average
6\% greater in other hippocampal layers if a heterogeneous tissue is
used instead of a homogeneous one
\cite{lopez-aguado_activity-dependent_2001, mcintyre_excitation_1999}.

For varying extracellular resistivity a numerical procedure referred to
as `the method of images' \cite{weber_1950} has been proposed as an extension to the point source equation. Although this method has been used in computations of extracellular
action potentials in a previous study \cite{gold_origin_2006}, its
practical use is limited to a rather low number of resistivity layers and
non-complex geometry.

The requirement of FEM to model passive current flow between neurons
was suggested elsewhere \cite{bedard_modeling_2004,
  bedard_model_2006}, but implementation issues and the lack of
adequate software tools may have precluded its usage in the past.

Extracellular resistivity could also contribute to the amplitude of
local field potentials (LFP), and in fact, in vitro LFP registered in
hippocampal slices are greater in the SP than in other layers
\cite{fisahn_cholinergic_1998}. Interestingly, we observed that
hippocampal slices in interface-type recording chambers (where slices
are not completely submerged in the buffer solution
\cite{inaba_volume-conducted_2006, hajos_maintaining_2009}) are more
than 60\% less conductive than in the chambers where slices are
submerged (unpublished results). This could help explaining why LFP
recorded in interface-type cambers are far greater than in
submerged-type counter parts \cite{leao_kv7/kcnq_2009,
  hajos_maintaining_2009}.

Here we show that the use of FEM software (COMSOL Multyphysics), with
an interface to Hodgkin and Huxley-type ODE model in Matlab, is a
powerful tool to verify macroscopic effects of neural fields.
However, this approach may fail to simulate small nuances of extracellular
interactions (e.g. ion channels apposing active compartments may
suffer more from the effect of passive current flow than channels in
the opposite side). Nonetheless, simulations using PDEs could increase
the level of realistic models of membrane dynamics. The idea of
translating membrane dynamics to PDE was proposed by Hodgkin and
Huxley themselves \cite{hodgkin_quantitative_1952}, and later pursued
by Kashef and Bellman \cite{kashef_solution_1974}. However,
implementing membrane dynamics in combination with the Maxwell
equation interface of the PDE simulator has been, to our experience,
quite cumbersome. For example, due to the relatively high
computational cost of the solution phase of the finite element method,
a representation of fully reconstructed morphologies with volumetric
cylindrical elements was not possible. The high amount of (small)
cylindrical elements required to accurately model complex dendritic
branches typically caused the meshing algorithm to break.

The software used in our simulations (COMSOL Multiphysics) has helped
to popularize the use of FEM in neuroscientific
problems \cite{zhang_modeling_2010, yousif_evaluating_2010}. However,
the software is geared towards industrial applications and the steep
learning curve associated to the adaptation of COMSOL to neuroscience
related problems may preclude its widespread usage by the neuroscience
community.  Possible combination of Neuron Simulation Environment (NSE) with a FEM simulator would offer great possibilities by allowing both realistic neural morphologies and extracellular space modeling. However, currently the existence of a direct interface between NSE and a FEM simulator has not been reported; future studies should focus on this task. Furthermore, hippocampal extracellular stimulation has been proposed as therapeutic strategy to control several disorders, including temporal lobe epilepsy \cite{stypulkowski_development_2011,luna-munguia_effects_2012}. Developing models that consider as many as possible physical tissue properties would help to improve extracellular stimulation in epilepsy patients. 
In summary, our work adds to the
recently published studies attempting to reveal important parameters
determining the strength of extracellular electric fields. Models that
do not use space and time-dependent differential equations when
modeling neuronal interactions may have failed to replicate the
changes in measured voltages caused by passive current flow in
heterogeneous extracellular tissue, especially when more than two
neurons are modeled simultaneously.

\section*{Acknowledgment}

This work was supported by Kjell and M\"arta Beijers Foundation and
the BMWF Marietta Blau stipend. PB and SE were supported by the
Swedish Research Council within the UPMARC Linnaeus center of
Excellence.


\bibliographystyle{IEEEtran}
\bibliography{manuscript}

\end{document}